\documentstyle[12pt,psfig]{article}

\def\thefootnote{\fnsymbol{footnote}}

\newcommand{\be}{\begin{equation}}      
\newcommand{\ee}{\end{equation}}        
\newcommand{\bear}{\begin{eqnarray}}    
\newcommand{\eear}{\end{eqnarray}}      
\newcommand{\beqstar}{\begin{eqnarray*}}        
\newcommand{\eeqstar}{\end{eqnarray*}}  

\newcommand{\gsim}{ \mathop{}_{\textstyle \sim}^{\textstyle >} }
\newcommand{\lsim}{ \mathop{}_{\textstyle \sim}^{\textstyle <} }

\begin{document}

%\markboth{H.-C. Cheng}{Universal Extra Dimensions at the $e^- e^-$ colliders}

%\catchline{}{}{}

\begin{titlepage}

\begin{minipage}[t]{3in}
\begin{flushleft}
%\today
June 3, 2002
\end{flushleft}
\end{minipage}
\hfill
\begin{minipage}[t]{3in}
\begin{flushright}
EFI-2002-86\\
\end{flushright}
\end{minipage}

\begin{center}
{\Large \bf Universal Extra Dimensions\\ at the $e^- e^-$ Colliders\footnote{
Talk presented at the 4th International Workshop on Electron-Electron
Interactions at TeV Energies, University of California, Santa Cruz, 
Dec.7-9, 2001}} 

\vskip .5in 

{\normalsize \bf Hsin-Chia Cheng}

\vskip .5in
 
{\small \it Enrico Fermi Institute, The University of Chicago, Chicago, 
  IL 60637, USA\footnote{Email: hcheng@theory.uchicago.edu}
}   
     
%\date{\today}

\vskip .5in

\begin{abstract}
{\normalsize  }
Universal Extra Dimensions (UEDs) with compactification radius near the TeV
scale provide interesting phenomenology at future colliders.
The collider signals of the first Kaluza-Klein (KK) level are very similar
to those of a supersymmetric model with a nearly degenerate
superpartner spectrum. The heavier first level KK states cascade decay
to the lightest KK particles (LKP), which is neutral and stable
because of KK-parity. The signatures involve missing energy and relatively
soft jets and leptons which can be difficult for detection.
The KK electron signal in $e^- e^-$ collisions is free from the
problematic two photon background therefore provides a unique
opportunity for a detailed studies of the KK electrons in the
Universal Extra Dimension scenario.
\end{abstract}

\vskip 1in
%PACS number: 

\end{center}
\end{titlepage}

\renewcommand{\thefootnote}{\arabic{footnote}}
\setcounter{footnote}{0}
\pagestyle{plain}

\section{Introduction}	%) A SECTION HEADING

Large extra dimensions
have recently attracted a lot of interest. They provide many new
theoretical ideas to explore questions in particle physics.
Most excitingly, they also predict signals which will be tested
at the upcoming collider experiments.

In this talk we focus on the scenario of Universal Extra Dimensions
(UEDs)~\cite{Appelquist:2001nn}. In UEDs, all standard model (SM)
fields propagates in extra dimensions of size $R \sim \mbox{TeV}^{-1}$.
There are many theoretical motivations for this scenario, such as
electroweak symmetry breaking~\cite{Arkani-Hamed:2000hv}, 
proton decay~\cite{Appelquist:2001mj},
the number of generations~\cite{Dobrescu:2001ae},
neutrino masses~\cite{Appelquist:2002ft}, etc.
The current experimental bounds allow Kaluza-Klein (KK) states in
UEDs to be as light as a few hundred 
GeV~\cite{Appelquist:2001nn,Agashe:2001xt,Appelquist:2001jz}.
Such light KK states can be copiously produced at the current or
future colliders with center of mass energies $\gsim$ TeV. 
It is therefore interesting to study their collider phenomenology,
and one might expect that it would be easy to discover these KK states.
However,
although their production cross section can be very large, 
their subsequent detection is
non-trivial because they decay nearly invisibly.
The phenomenology of UEDs shows interesting parallels to
supersymmetry. Every Standard Model field has KK partners. The
lowest level KK partners carry a conserved quantum number,
KK parity, which guarantees that the lightest KK particle (LKP) is stable.
Heavier KK modes cascade decay to the LKP by emitting soft
Standard Model particles. The LKP escapes detection, resulting
in missing energy signals. 

Given that the signals are very similar to supersymmetry (SUSY),
it is important to distinguish them by studying the properties
of the new states produced in colliders in details. Linear colliders
often provide a clean environment for such kind of studies.
However, in $e^+ e^-$ collisions, the signals will be plagued
by the huge two photon backgrounds due to the softness of the observable
decay products. On the other hand, the $e^- e^-$ mode, as we will see, 
stands as a unique opportunity for a detailed study of the KK excitations
of the electrons. This talk was based on work in collaboration with
Konstantin T. Matchev and Martin Schmaltz~\cite{Cheng:2002iz,Cheng:2002ab}.

\section{Minimal Universal Extra Dimensions}

Let us first describe the model.
The simplest UED scenario has all of the Standard Model fields
(no supersymmetry) propagating in a single extra dimension.
In 4+1 dimensions, the fermions 
[$Q_i,\, u_i, \, d_i,\, L_i,\, e_i, \, i=1,2,3$, where upper (lower) case 
letters represent $SU(2)$ doublets (singlets)] are four-component and contain
both chiralities when reduced to 3+1 dimensions. To produce a chiral
4d spectrum, we compactify the extra dimension on an $S_1/Z_2$ orbifold.
Fields which are odd under the $Z_2$ orbifold symmetry
do not have zero modes, hence the unwanted fields (zero modes of fermions with
the wrong chiralities and the 5th component of the gauge fields)
can be projected out. The remaining zero modes are just the Standard Model
particles in 3+1 dimensions.

The full Lagrangian of the theory comprises both
bulk and boundary interactions.
Gauge and Yukawa couplings and the Higgs potential are contained in the
bulk Lagrangian in one-to-one correspondence with
the couplings of the Standard Model. The boundary
Lagrangian interactions are localized at the orbifold
fixed points and do not respect five dimensional Lorentz invariance.

Ignoring the localized terms for the moment, the tree-level mass of 
the $n$-th KK
mode is
\begin{equation}
\label{tree-spectrum}
m_n^2= \frac{n^2}{R^2} + m_0^2,
\end{equation}
where $R$ is the radius of the compact dimension, and $m_0$ is the zero
mode mass. The spectrum at each KK level is highly degenerate except
for particles with large zero mode masses ($t,\,  W,\, Z,\, h$).
The bulk interactions preserve the 5th dimensional momentum
(KK number). The corresponding coupling constants among KK modes 
are simply equal to
the SM couplings (up to normalization factors such as $\sqrt{2}$).
The Feynman rules for the KK modes can easily be derived 
(e.g., see Ref.~\cite{Macesanu:2002db,Dicus:2000hm}
).

In contrast, the coefficients of the boundary terms are not
fixed by Standard Model couplings and correspond to new
free parameters. In fact, they are renormalized by the bulk interactions
and hence are scale dependent~\cite{Georgi:2001ks,Cheng:2002iz}.
One might worry
that this implies that all predictive power is lost.
However, since the wave functions of Standard Model fields and KK modes are
spread out over the extra dimension and the new couplings only exist on the
boundaries, their effects are volume suppressed. We can get an
estimate for the size of these volume suppressed corrections with naive
dimensional analysis by assuming strong coupling at the cut-off.
The result is that the mass shifts to KK modes from boundary terms
are numerically equal to corrections from loops
$\delta m_n^2/m_n^2 \sim g^2/16 \pi^2$.

We will assume a symmetry which exchanges the two orbifold fixed points.
The boundary terms on the two fixed points are equal under this symmetry.
It is anomaly free and preserved by the radiative corrections of the bulk 
interactions.
Most relevant to the phenomenology are localized kinetic
terms for the SM fields, such as
\begin{equation}
\label{newops}
\frac{\delta(x_5)+\delta(x_5-\pi R)}{\Lambda} 
\left[
G_4 (F_{\mu\nu})^2 
+ F_4 \overline \Psi i \slash\!\!\!\!D \Psi + 
F_5 \overline \Psi \gamma_5 \partial_5 \Psi
\right], 
\end{equation}
where the dimensionless coefficients $G_4$ and $F_i$ are arbitrary and
not universal for the different Standard Model
fields. These terms are important phenomenologically for several 
reasons: ({\it i}) they split the near-degeneracy of KK modes at each level,
({\it ii})
they break KK number conservation down to a KK parity
(due to the symmetry between the two fixed points,)
under which modes with odd KK numbers are charged,
({\it iii}) they may introduce new flavor violations.

For simplicity and definiteness, we will concentrate on the 
Minimal Universal Extra
Dimensions (MUEDs)~\cite{Cheng:2002ab}, where all boundary terms
are assumed to vanish at some cutoff scale $\Lambda > R^{-1}$.
At a lower scale, they will be generated by radiative corrections from 
the bulk interactions which are calculable. This is analogous to
the case of the Minimal Supersymmetric Standard Model (MSSM) where
one has to choose a set of soft supersymmetry breaking parameters at some
high scale in order to study its phenomenology. In some sense,
our choice of the boundary terms may be viewed as analogous
to the simplest minimal supergravity boundary condition --- universal
scalar and gaugino masses.

Given this choice of the boundary condition, we can calculate the
corrections to the KK spectrum, which depends on the boundary terms
at low scales. Since the corrections are small one can use the one-loop
leading log approximation. In addition to the boundary terms, the
KK spectrum also receives corrections from pure bulk effects.
Loops wrapping around the compact extra dimension with nonzero net
winding numbers break the 5-dimensional Lorentz invariance and
therefore also correct the KK mass formula \ref{tree-spectrum}.
The bulk corrections are finite and exactly calculable due to
their non-local nature. All these corrections, including both
bulk and boundary contributions, were computed at one-loop in
Ref.~\cite{Cheng:2002iz}.

\begin{figure}%[tb]
\centerline{\psfig{file=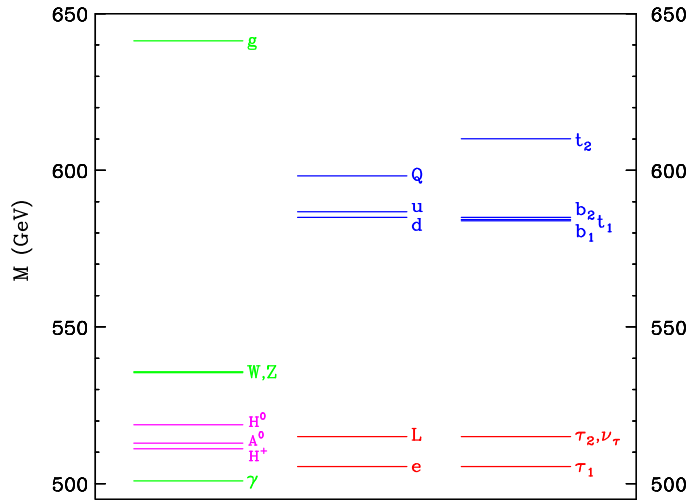,width=9cm}}
\caption{\label{fig:spectrum} { One-loop corrected mass spectrum of
the first KK level in MUEDs for $R^{-1}=500$ GeV, $\Lambda R = 20$
and $m_h=120$ GeV.}}
\end{figure}
An example of the spectrum for the first level KK modes is shown in
Fig.~\ref{fig:spectrum}.
Typically, the corrections for KK modes with strong interactions are 
$> 10\%$ while those for states with only electroweak interactions are a few
percent. We find that the corrections to the masses are such that 
$m_{g_n}>m_{Q_n}>m_{q_n}>m_{W_n}\sim m_{Z_n}>m_{L_n}>m_{\ell_n}>m_{\gamma_n}$.
The lightest KK particle $\gamma_1$, is a mixture of the first KK 
mode $B_1$ of the $U(1)_Y$
gauge boson $B$ and the first KK mode $W^0_1$ of the $SU(2)_W$ $W^3$ 
gauge boson.
We will usually denote this state by $\gamma_1$. However, note that
the corresponding ``Weinberg'' angle $\theta_1$ is much smaller than
the Weinberg angle $\theta_W$ of the Standard Model~\cite{Cheng:2002iz},
so that the $\gamma_1$ LKP is mostly $B_1$ and $Z_1$ is mostly $W^0_1$.
The mass splittings among the level 1 KK modes are large enough 
for the prompt decay of a heavier level 1 KK mode to a lighter 
level 1 KK mode. But since the spectrum
is still quite degenerate, the ordinary SM particles emitted from
these decays will be soft, posing a challenge for collider searches.
 
The terms localized at the orbifold fixed points also violate the KK number 
by even units. However, assuming that no explicit KK-parity violating 
effects are put in by hand, KK parity remains an exact symmetry.
The boundary terms allow higher ($n>1$) KK modes to decay to lower KK modes,
and even level states can be singly produced (with smaller cross sections
because the boundary couplings are volume suppressed). Thus
KK number violating boundary terms are important for higher KK mode 
searches.

\section{Collider Phenomenology}

\begin{figure}%[tb]
\centerline{\psfig{file=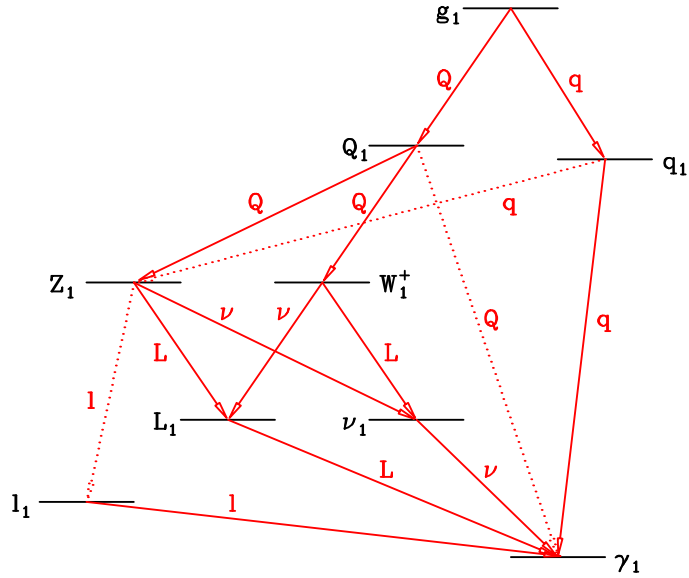,width=9cm}}
\caption[]{ Qualitative sketch of the level 1
KK spectroscopy depicting the dominant (solid) and rare (dotted) 
transitions and the resulting decay product.
\label{fig:transitions}}
\end{figure}

Once the radiative corrections are included, the KK mass degeneracy at 
each level is lifted and the KK modes decay promptly. The collider
phenomenology of the first KK level is therefore very similar to
a supersymmetric scenario in which the superpartners are relatively
close in mass --- all squeezed within a mass window of 100-200 GeV
(depending on the exact value of $R$). Each level 1 KK particle has
an exact analogue in supersymmetry: $B_1\leftrightarrow$ bino,
$g_1\leftrightarrow$ gluino, $Q_1(q_1)\leftrightarrow$ left-handed
(right-handed) squark, etc. The decay cascades of the level 1 KK modes
will terminate in the $\gamma_1$ LKP (Fig.~\ref{fig:transitions}).
Just like the neutralino LSP is stable in $R$-parity conserving
supersymmetry, the $\gamma_1$ LKP in MUEDs is stable due to KK parity 
conservation and its production at colliders results in generic
missing energy signals. 

At hadron colliders, the largest cross sections for KK states come from
strongly interacting KK quarks and gluons because of the approximately
degenerate KK spectrum. The level 1 KK states have to be pair produced
due to KK parity conservation. Their cross sections at the Tevatron
Run II and the LHC can be found in 
Ref.~\cite{Macesanu:2002db} and~\cite{Rizzo:2001sd}. In MUEDs, the 
heaviest KK state at level 1 is the KK gluon $g_1$. It decays
to a KK quark and an ordinary quark. The $SU(2)$ singlet KK quarks
will decay to $\gamma_1$ directly. The signature is jets$+\not \!\!E_T$,
similar to the traditional squark and gluino searches~\cite{squarks}.
Using the existing studies of the analogous SUSY studies, one might
expect that Run II can probe $R^{-1}\sim 300$ GeV while the LHC
reach for $R^{-1}$ can be
$\sim 1$ TeV~\cite{Abel:2000vs,Bityukov:1999am}.
However, the jets will be relatively soft in the case of MUEDs due to the
approximate degeneracy of the  KK initial and final states, and the measured
missing energy is correlated with the energy of the soft recoiling jets
hence also small, even though the total missing mass is large.
They may cause more difficulties in triggering and separation of
backgrounds. Further studies in the context of MUEDs are needed to
obtain a better estimate of the reaches with the jetty signatures.
The $SU(2)$ doublet KK quarks will mostly decay to KK $W$ and $Z$ bosons,
then subsequently decay to KK leptons and $\gamma_1$. They give rise to
the much cleaner signatures with multi-leptons. Using the very clean 
4 lepton channel in Ref.~\cite{Cheng:2002ab} we estimate that Run IIb can go slightly
beyond the current indirect bounds ($R^{-1} > 300$ GeV) from precision
data, and LHC can extend the reach to $R^{-1} \sim 1.5$ TeV.

Although LHC will be able to extend the reach much beyond the current
bounds, to identify the extra-dimensional nature of the new physics
once some signals are found will be rather challenging. All signals from 
level 1 KK states look very much like supersymmetry --- all SM particles
have ``partners'' with similar couplings. Of course, the spins of
these SM particle partners are different in SUSY and UEDs, but this
difference will most likely escape detection at a hadron collider.
In addition, all the observed jets, leptons, and $\not \!\! E_T$ are
relatively soft. It would be difficult to know exactly the mass scale
of these KK states.
Additional experimental information will be needed to identify which
scenario of new physics is realized and the mass scale associated with it. 

\begin{figure}%[tb]
\centerline{\psfig{file=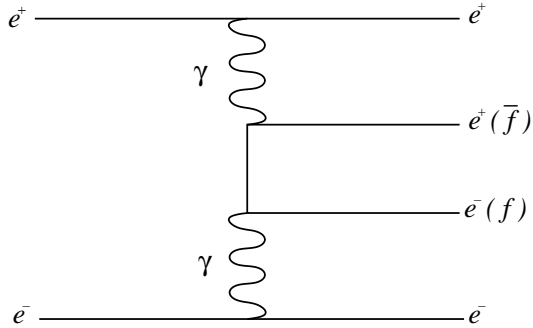,width=8cm}}
\caption[]{ The backgrounds from the 2 photon processes.
\label{fig:2photon}}
\end{figure}

Linear colliders provide a clean environment for precise studies of particle
properties. Taking the slepton production in supersymmetry as an example,
one can get a precise measurement of the slepton mass and the LSP mass from
the end points of the final state lepton energy 
distribution~\cite{Tsukamoto:1993gt,Abe:2001np}.
The flat energy distribution between the end points will also tell
the scalar nature of the sleptons produced. Therefore, a linear collider
can be a great tool to distinguish UEDs from SUSY.
Given the bounds on the masses of the level 1 KK modes, the required energy
for pair producing them is probably too high for the first stage of the
next generation linear colliders, but may be reachable with energy upgrades
or at CLIC. However, in $e^+ e^-$ collisions there are
huge backgrounds from the two photon processes for very soft final state
leptons or jet (Fig.~\ref{fig:2photon}). Soft leptons or jets are produced 
by the collisions of the two soft photons from the initial state radiation,
while the initial $e^+$ and $e^-$ go down the beam pipe, resulting in
large missing energies.
The two photon background for slepton search at a 500 GeV linear collider
is studied in details in Ref.~\cite{colorado}. 
It was shown that for very soft final
state leptons ($\lsim 10\% \times E_{\rm beam}$), the two photon background is 
still several order magnitude larger than the signal even after the cuts.
In the case of MUEDs, the mass splittings between the KK leptons and
$\gamma_1$ are only about 1--3\%. Therefore the leptons coming from decays
of the KK leptons will be completely buried in the two photon background.
KK quarks receive larger mass corrections and the jets coming from their
decay would be harder. Their detection in the $e^+ e^-$ collisions
is more promising.

On the other hand, only KK electrons can be produced in the $e^- e^-$
collisions (if there is no flavor violation from the boundary terms).
The signal will be two soft electrons of the {\it same} charge plus
large missing energy, while the two photon processes produce fermions
of the opposite charges. By charge identification one can remove the
problematic two photon background. Therefore, the $e^- e^-$ mode of a
linear collider stands as a unique opportunity for KK electron studies.
All particle properties of the KK electrons should be able to be 
measured unambiguously, providing valuable information for identifying 
the corresponding new physics. It would also be useful to study possible
lepton flavor violations if they are present as in the case of 
supersymmetry~\cite{Arkani-Hamed:1996au}.

\section{Conclusions}

Universal extra dimensions with compactification radius near the TeV scale
promise exciting phenomenology for future colliders. All Standard Model
particles have KK partners which can be produced with enormous cross
sections at the LHC. However, the subsequent decays produce very soft
Standard Model particles which can be difficult to see above the backgrounds,
hence their detection and identification are rather non-trivial.
The $e^+ e^-$ collider has one more piece of information as the total
missing energy can be measured, but also suffers from the huge two photon
background if the SM particles from the decays of the KK states are too
soft. The $e^- e^-$ collider provide a unique clean environment for
the detailed studies of the KK electrons. As different possible new
physics scenarios often give rise to complicated but sometimes similar
collider phenomenologies, it is essential to have as many tools as possible
to provide enough experimental information for our understanding
of any new physics.

\section*{Acknowledgements}
I would like to thank K.~T. Matchev and M.~Schmaltz for collaboration
on this work. I also thank N.~Arkani-Hamed, M.~Chertok, A.~Cohen, B.~Dobrescu
and B.~Schumm for useful discussions.
This work is supported by the Department of Energy grant DE-FG02-90ER-40560.  
%M.S. is supported in part by the Department of Energy under grant
%number DE-FG02-91ER-40676.

%\appendix

%\section{Appendices}

\end{document}